\documentclass[aps, prl, 10pt, twoside, twocolumn, superscriptaddress, showkeys, floatfix]{revtex4-1}
\usepackage{wrapfig}
\usepackage{upgreek}
\usepackage{latexsym,amssymb}
\usepackage[intlimits]{amsmath}
\usepackage{graphicx}
\usepackage{float}
\usepackage{physics}
\usepackage{color}
\usepackage{natbib}
\usepackage[hidelinks]{hyperref}

\begin{document}
\title{Molecular quantum spin network controlled by a single qubit}
\author{Lukas Schlipf}
\affiliation{Max Planck Institute for Solid State Research, Heisenbergstra\ss e 1, 70569 Stuttgart, Germany}
\affiliation{3.\ Physikalisches Institut, Universit\"at Stuttgart, Pfaffenwaldring 57, 70569 Stuttgart, Germany}
\author{Thomas Oeckinghaus}
\affiliation{3.\ Physikalisches Institut, Universit\"at Stuttgart, Pfaffenwaldring 57, 70569 Stuttgart, Germany}
\author{Kebiao Xu}
\affiliation{Max Planck Institute for Solid State Research, Heisenbergstra\ss e 1, 70569 Stuttgart, Germany}
\affiliation{3.\ Physikalisches Institut, Universit\"at Stuttgart, Pfaffenwaldring 57, 70569 Stuttgart, Germany}
\affiliation{CAS Key Laboratory of Microscale Magnetic Resonance and Department of Modern Physics, University of Science and Technology of China, Hefei 230026, China}
\author{Durga~Bhaktavatsala~Rao~Dasari}
\affiliation{Max Planck Institute for Solid State Research, Heisenbergstra\ss e 1, 70569 Stuttgart, Germany}
\affiliation{3.\ Physikalisches Institut, Universit\"at Stuttgart, Pfaffenwaldring 57, 70569 Stuttgart, Germany}
\author{Andrea Zappe}
\affiliation{3.\ Physikalisches Institut, Universit\"at Stuttgart, Pfaffenwaldring 57, 70569 Stuttgart, Germany}
\author{Felipe F\'avaro de Oliveira}
\affiliation{3.\ Physikalisches Institut, Universit\"at Stuttgart, Pfaffenwaldring 57, 70569 Stuttgart, Germany}
\author{Bastian Kern}
\affiliation{Max Planck Institute for Solid State Research, Heisenbergstra\ss e 1, 70569 Stuttgart, Germany}
\author{Mykhailo~Azarkh}
\affiliation{Department of Chemistry, Zukunftskolleg, and Konstanz Research School Chemical Biology, University of Konstanz, Universit\"atsstra\ss e 10, 78457 Konstanz, Germany}
\author{Malte Drescher}
\affiliation{Department of Chemistry, Zukunftskolleg, and Konstanz Research School Chemical Biology, University of Konstanz, Universit\"atsstra\ss e 10, 78457 Konstanz, Germany}
\author{Markus Ternes}
\affiliation{Max Planck Institute for Solid State Research, Heisenbergstra\ss e 1, 70569 Stuttgart, Germany}
\author{Klaus Kern}
\affiliation{Max Planck Institute for Solid State Research, Heisenbergstra\ss e 1, 70569 Stuttgart, Germany}
\affiliation{Institut de Physique, \'Ecole Polytechnique F\'ed\'erale de Lausanne, 1015 Lausanne, Switzerland}
\author{J\"org Wrachtrup}
\affiliation{Max Planck Institute for Solid State Research, Heisenbergstra\ss e 1, 70569 Stuttgart, Germany}
\affiliation{3.\ Physikalisches Institut, Universit\"at Stuttgart, Pfaffenwaldring 57, 70569 Stuttgart, Germany}
\author{Amit Finkler}
\email[{Author to whom correspondence should be addressed.
   Email:~}]{a.finkler@physik.uni-stuttgart.de}
\affiliation{3.\ Physikalisches Institut, Universit\"at Stuttgart, Pfaffenwaldring 57, 70569 Stuttgart, Germany}


\begin{abstract}
Scalable quantum technologies will require an unprecedented combination of precision and complexity for designing stable structures of well-controllable quantum systems. It is a challenging task to find a suitable elementary building block, of which a quantum network can be comprised in a scalable way. Here we present the working principle of such a basic unit, engineered using molecular chemistry, whose control and readout are executed using a nitrogen vacancy (NV) center in diamond. The basic unit we investigate is a synthetic polyproline with electron spins localized on attached molecular sidegroups separated by a few nanometers. We demonstrate the readout and coherent manipulation of very few ($\leq 6 $) of these $S=1/2$ electronic spin systems and access their direct dipolar coupling tensor. Our results show, that it is feasible to use spin-labeled peptides as a resource for a molecular-qubit based network, while at the same time providing simple optical readout of single quantum states through NV-magnetometry. This work lays the foundation for building arbitrary quantum networks using well-established chemistry methods, which has many applications ranging from mapping distances in single molecules to quantum information processing.
\end{abstract}

\maketitle
\date{\today}

\section*{Introduction}
Coherent control over a many-qubit system, as well as readout of few or single elementary qubits within such a system, is key for quantum information processing techniques. Among the different realizations of qubits, electron and nuclear spins in solids show unprecedented coherence times up to six hours~\cite{Zhong2015}, can be coherently controlled at GHz rates~\cite{Jelezko2004} and read out optically~\cite{Gruber1997} as well as electronically~\cite{Bourgeois2015}. Yet scaling of e.g.\ spin systems to larger regular arrays is a daunting technical task as typical distances among electron spins for them to couple via magnetic dipole interaction (below 30 nm~\cite{Dolde2013}) are currently out of reach for reliable top-down nanotechnology. Programmable molecular structure, however, can very well cover these length scales~\cite{Rothemund2006, Barth2005}, for example using sequence controlled self assembly of peptides on surfaces~\cite{Abb2016}. Such systems show the high degree of regularity required for complex patterns, built up from simple interacting quantum systems. In addition, their flexible chemistry allows to bring them into virtually any shape such that their design can adopt various spin lattices in two and three dimensions \cite{Gopinath2016, Luo2016}. Here we demonstrate measurement of a small basic unit cell for the goal of providing both scalability as well as a mechanism for quantum information readout, comprising two electron spins positioned on a molecule. Probing the coupling between the two spins and ultimately driving them in a coherent fashion constitutes an essential building block for a molecular spin network.

Quantum spin networks are known to integrate and scale-up quantum registers involving many qubits. Combined with local control in implementing quantum logic gates, spin chains can greatly facilitate large-scale quantum information processing \cite{Bose2003, DiFranco2008, Ajoy2013b}. In addition to the naturally available spin chains in materials \cite{Sahling2015}, artifical spin chains are realized with quantum dots \cite{McNeil2011}, chains of magnetic atoms \citep{Loth2012}, Josephson junction arrays \cite{Geerligs1989}, and trapped ions \cite{Debnath2016}. While the spin-spin distances in materials range in the $ 10\ \textrm{nm}$ regime, offering scalability and high coupling strength among nearest neighbors, they are less controllable even at the few-spin level. On the other hand, artificially formed spin chains are more controllable, allowing them for example to be used as simulators for understanding energy transport in complex networks \cite{Park2016}. In the intermediate regime of length scales in the nm-region, molecular chemistry offers stable scaffolds to build scalable spin networks. These systems have mostly been studied in the context of magnetic resonance for distance measurements among the spins \citep{Jeschke2012}, yet typically in the ensemble-limit of around $10^9$ spins due to sensor limitations. It is therefore necessary to find a suitable sensor system, that can read out quantum information from such a spin network on the local scale of few spins in a non-invasive fashion.
\begin{figure*}[htb!]
 \includegraphics[width=0.9\textwidth]{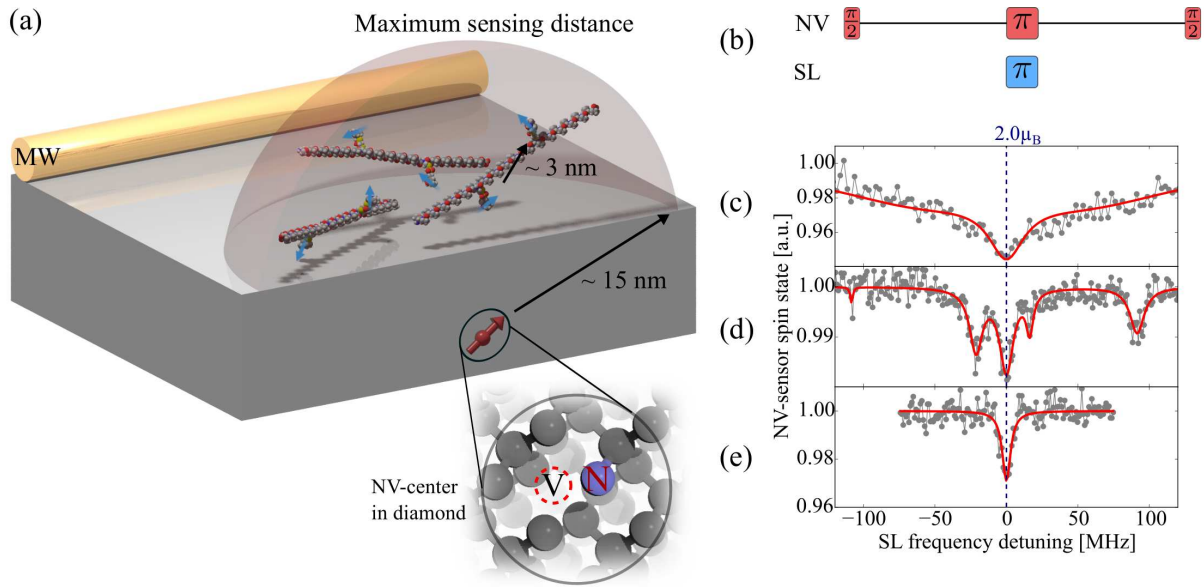}
 \caption{Electron spin spectra measured at $T = 4.2\ \mathrm{K}$ and UHV; (a) Experimental sketch: The shallow NV-center spin couples to doubly spin-labeled polyprolines. Coherent control over all participating spins is provided by a close-by ($\sim 50\ \upmu \mathrm{m}$) microwave wire (MW). The edge of the depicted sphere is the maximum sensing radius for a single electron spin to be detected, delimited by the NV-centers' $T_2$-time. The intra-peptide coupling is almost two orders of magnitude larger than any of the other couplings in the few peptide case, especially the inter-peptide coupling. (b) Pulse sequence used to probe the spectrum of external network spins. Red pulses address the NV-center spin, while the frequency of the blue pulse is swept to address network spin labels (SL) (c) Measured spectrum on a diamond membrane coated exclusively with spin-labeled peptides. The fit of the spectrum is a numerical simulation over random orientations of 50 MTSSL spin labels including decoherence effects due to inter-peptide couplings. (d) Same as (c), but the labeled polyprolines were diluted with unlabeled ones at a ratio of 1:10. Distinct hyperfine couplings are clearly visible. (e) Measurement of a diamond dangling bond dark spin on a cleaned membrane for comparison.}
 \label{fig:fig1}
\end{figure*}

Over the past few years, a novel atom-sized defect in the form of the nitrogen vacancy (NV) center in diamond has revealed the ability to perform magnetic resonance studies on very few spins in a nanoscale volume \cite{Balasubramanian2008, Taylor2008}, where its quantum nature is harnessed for magnetometry using optical means. Quantum-limited sensitivity has been achieved using the NV-center in detecting single electron \cite{Grinolds2014, Shi2015} and nuclear spins \cite{Sushkov2014, Mueller2014} external to the diamond lattice. Its long coherence time \cite{Balasubramanian2009}, even in close proximity to the surface, makes it possible to sense a single electron spin at a distance more than 30 nm from the NV-center \cite{Grinolds2013}. This unique atomic-sized defect serves here as a probe for a network's unit cell, reading it out with high fidelity using confocal microscopy, while microwave pulses facilitate manipulation of the sensor and network spins.

\section*{Experiment and Theory}
We choose a synthetic, proline(P)-rich peptide as a spacer model system, H-AP$_{10}$CP$_{10}$CP$_{10}$-NH$_{2}$, containing two cysteine (C) amino-acids at specific positions, to which various spin labels can be easily attached via site directed spin labeling (SDSL). Due to the relatively high stiffness of the proline-backbone \cite{Schuler2005, Ungar1973}, the electron spins attached to the polyproline (network spins) are well-localized, making them resilient to phonon-induced decoherence, even at finite temperatures. These network electron spins are locally probed and controlled by their dipolar coupling to the closeby electron spin located on the NV-center in diamond (probe spin). To enable this coupling, we use shallow NV-centers close to the diamond surface ($< 10$ nm), on which we deposit the spin labeled peptides (Fig.\,\ref{fig:fig1}a). The higher gyromagnetic ratio of the network electron spins in comparison to nuclear spins makes it more feasible to drive coherent oscillations at high frequencies and detect their response \cite{Grotz2011}.

\begin{figure*}[bt!]
 \includegraphics[width=0.75\textwidth]{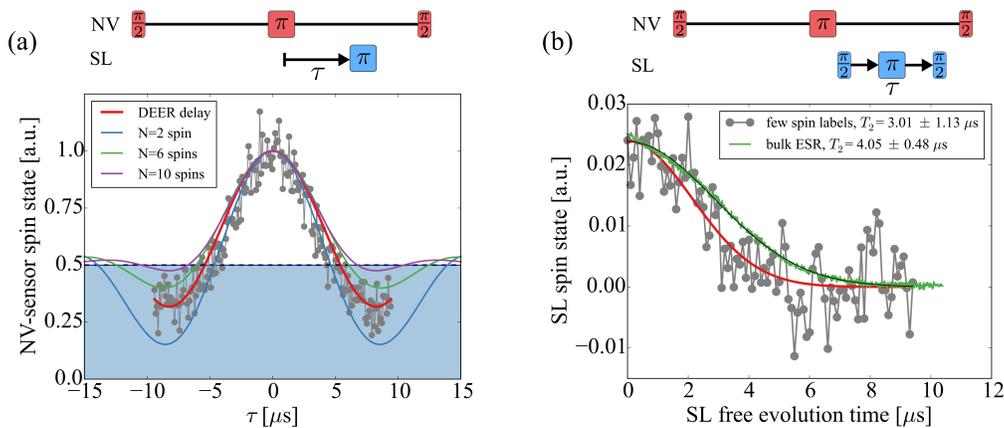}
 \caption{(a) Coherent oscillation due to dipolar couplings between sensor and network spins, clearly visible by contrast inversion (blue shaded area) of the NV probe spin state which suggests a low number of participating network spins. The measurement data is accompanied by simulations of coupling strength oscillations caused by $N$ = 2, 6, and 10 randomly distributed spins above the diamond surface. We would like to stress that it is difficult to estimate the precise number of spins here and in fact the relative strength of the overshoot below the NV-sensor mixed state (0.5) heavily depends on the couplings (positions) of the participating network spins. (b) Hahn echo measurement on the external network spins directly probing their $T_{2}$ time via the readable signal from the probe spin (red). The green curve was measured on a pulsed EPR spectrometer at 10 K in a frozen solution of water/glycerol (4:1) using the same peptides. Low concentration guarantees a proper separation between peptides and the agreement of both coherence times points towards comparable SDSL peptide-peptide distances in both samples.}
 \label{fig:fig2}
\end{figure*}

The sensor spin, viz.\ the NV-center, is a ground state paramagnetic $S = 1$ spin-system, which can be optically polarized with near unit efficiency to the non-magnetic ($m_z = 0$) ground state of the spin. Similarly, its spin-state projection on its quantization ($z$-)axis can also be read-out optically with high fidelity \cite{Gruber1997}. Due to a relatively large zero-field splitting of the NV-center, its spin state manipulation by microwave fields can be quite different from the network electron spins \cite{Jelezko2004}, enabling independent control. To efficiently characterize the spin network utilizing NV-center spin state readout, our first goal is to unravel the various couplings involved in the dynamics. For this we use double electron-electron spin resonance techniques (see Methods) as detailed in the following.

Here we choose the common methanethiosulfonate spin label (MTSSL) as a network spin which will lead to three distinct hyperfine peaks in the spectroscopy due to hyperfine coupling of its electronic $\sigma = 1/2$ spin with a $^{14}\mathrm{N}$ nuclear spin (total spin quantum number $\textrm{I}=1$) \citep{Shi2015}. Initially, we cover the NV-center containing diamond \textit{only} with MTSSL bearing peptides (see Methods). Using this sample preparation, almost every NV-center shows a strong electron spin signal peak (see SI). Although it is likely that $\geq 50$ spins are detected, spectra at ambient conditions showing two distinct sidepeaks can be found. This behavior is attributed to motional narrowing of the different inhomogeneous spectra, since peptides were not immobilized on the diamond surface and hence molecular tumbling can be significant \cite{Molaei2014}.

One well-known problem in optical excitation of organic molecules is photo-bleaching, i.e.\ chemical reactions and the accompanying structural and electronical changes induced by supplying energy in the form of photons. Since typically around $1\ \mathrm{mW}$ of power is focused on a diffraction-limited spot to saturate the NV-center fluorescence, the electron spins within dipolar coupling range to the NV-center also experience a power density of usually $10^{9}\ \mathrm{W}\cdot\mathrm{cm}^{-3}$, which can often lead to a change or loss of the electronic spin properties in the molecule. Indeed, we observe a bleaching of the MTSSL electron spin contrast on an hour timescale in almost all cases, preventing more elaborate experiments. To extend this timescale, often lowering the temperature of the system under investigation \cite{Moerner1999} as well as stripping the environment of oxygen to prevent photo-catalytic reactions is of great benefit. Consequently, all of the following experiments were conducted at liquid helium temperatures (4.2 K) and in an ultra high vacuum environment (UHV, $\textrm{p} < 10^{-8} \textrm{ Pa}$) \cite{SchaeferNolte2014}, where we do not observe photobleaching of the organic network spins, demonstrated by an excellent signal stability.

Upon cooling the dense network spin sample we described above, the previously observed narrow lines uniformly show significant broadening (Fig.\,\ref{fig:fig1}c). This can be explained by the freezing-out of any network spin motion on the timescale of the measurement and the resulting loss of motional narrowing. In this case the resulting spectrum can be well described by a sum over all possible orientations \cite{supplement}. The high network spin density has a great effect on the coherence of the electronic network spins due to electron-electron flip flop processes, while at the same time the magnitude of sensor-network couplings makes it impossible to access single network spin states. We therefore reduced the sensor-network coupling to only a few spins.

To this end we diamagnetically dilute the peptide stock solution using non-labeled peptides with a ratio of 1:10 before deposition on the diamond surface, increasing the inter-peptide spin-spin distance in the resulting homogeneously mixed molecular film. Subsequently, we observe spectra on NV-centers with narrow peaks at low temperature (Fig.\,\ref{fig:fig1}d), suggesting low spin-spin interaction between the network spins and thus probe-to-network coupling with only a few participating spins as illustratively depicted in Fig.\,\ref{fig:fig1}a. Due to the rather high concentration of spin-labeled molecules, the ratio of NV-centers exhibiting broad peaks to narrow peaks was approximately 40:1. We select an NV-center showing four distinct sidepeaks (Fig.\,\ref{fig:fig1}d) for all following measurements, possibly caused by pairs of electron spins at different angles and positions as can be found from a few peptides.

Fixing the $\pi$-pulse transition frequency at any of the hyperfine peaks we vary the application timing $\tau$ of the network spin pulse within the sequence (Fig.\,\ref{fig:fig2}a), or equivalently vary the total length of the sequence (not shown). With this and various sensor-to-network couplings $F^k$ involved, the resulting population difference among the probe spin states $C_{\textrm{NV}}$ oscillates as
\begin{equation*}
 C_{\textrm{NV}}(\tau) = \frac{1}{2} \left[1 + \prod^M_{k}{\cos\left( F^k \tau \right)}\right].
\end{equation*}
From the above equation one can see that population inversion of the probe i.e., $C_S < 0.5$ occurs for a single or few spins and slowly vanishes (i.e., $C_S \ge 0.5$) with increasing number of spins in the sensing volume of the NV-center probe spin. 

Consequently, to gain information on the number of participating network spins we apply the above described sequence, showing a coherent oscillatory behavior (Fig.\,\ref{fig:fig2}a) exclusively for the narrow peak case as on the previously selected sensor spin. From the observed population inversion in the acquired measurement data (i.e.\ $C_S < 0.5$) and from the density of the spin labels deposited on the diamond we estimate a small number of spins ($\lesssim 6$) coupling to this sensor, confirming the few spin limit. For the broad spectra, as for example in Fig.\,\ref{fig:fig1}c, the measurement likewise shows a pure decay to the sensor spin's thermal state ($C_S = 0.5$).

\begin{figure}[hbt!]
 \includegraphics[width=0.8\columnwidth]{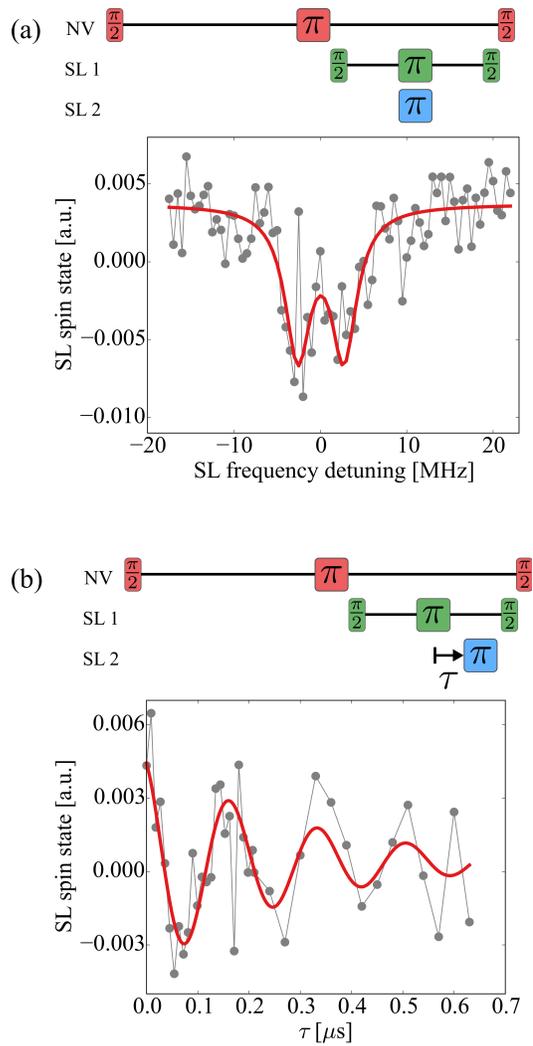}
  \caption{(a) Measurement of a triple resonance frequency sweep few network spins (Fig.\,\ref{fig:fig1}d, Fig.\,\ref{fig:fig2}) with a $\pi$-pulse width of $3.4$ MHz. The resulting spectrum can be fitted with two Lorentzians (red), as is expected from a pronounced $J$-coupling between network spins. (b) Delay measurement on a the same environment, where we fix the excitation frequency of SL 1 at the network Larmor frequency and the SL 2 sweeping $\pi$-pulse frequency 3 MHz apart from it. The red curve is a fit using a decaying cosine. The oscillations reveal the very same coupling strength as the frequency sweep in (a).}
 \label{fig:fig3}
\end{figure}

In a similar measurement we determine the coherence time of the network spins by the pulse sequence shown in Fig.\,\ref{fig:fig2}b. The measured phase coherence time of $T_2 = (3.0 \pm 1.1)\ \upmu\mathrm{s}$ (Fig.\,\ref{fig:fig2}b) is in very good agreement with standard low temperature pulsed ESR measurements on diluted peptides in solution \cite{supplement}. Although the two measurements are performed in different peptide environments and are difficult to compare, this again confirms that spin-labeled peptides are well-separated. Another indication for this is the Gaussian decay of coherence, a sign for a proton flip-flop limited decoherence \cite{Abragam1983} with no electron bath contribution. In addition to the ability to perform and read-out single spin (qubit) gates, the coherent dipolar coupling within the network allows for performing conditional two-spin gates. For this however, the extraction of the $J$-coupling among these spins is required, which is not readily possible with the so-far described pulse sequence. 

\begin{figure*}[hbt!]
 \includegraphics[width=0.85\textwidth]{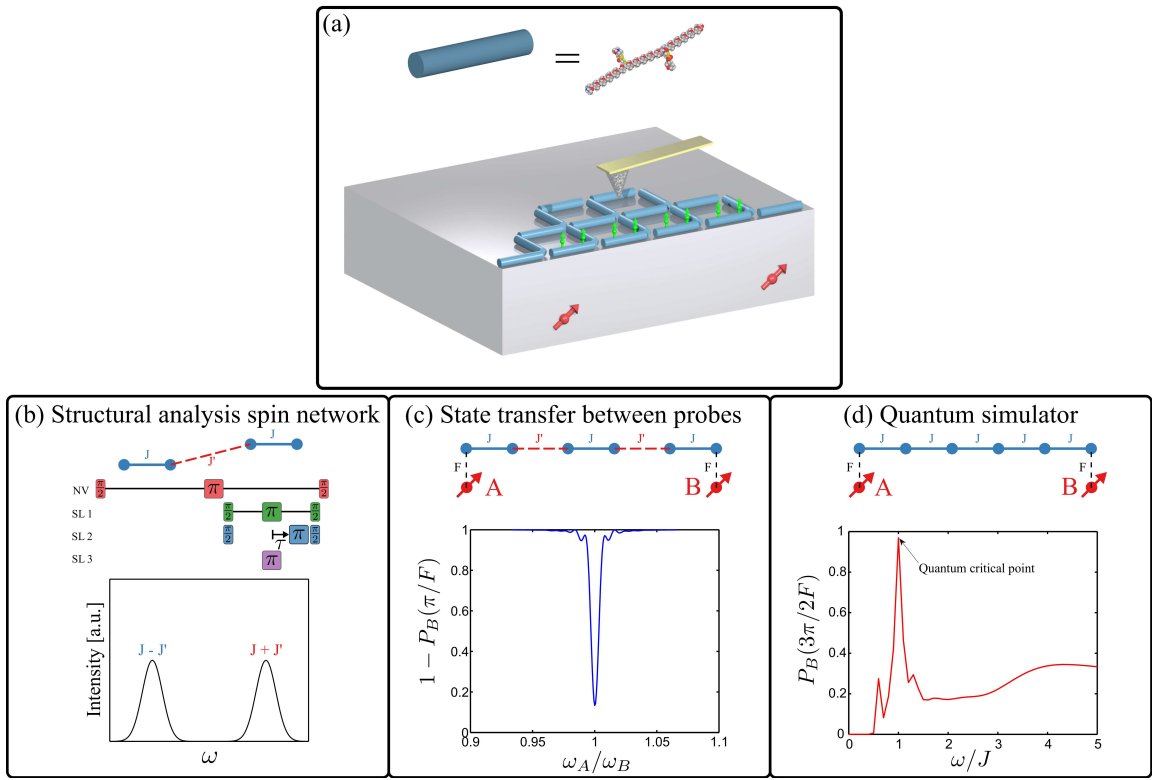}
  \caption{(a) Schematic representation of the envisaged probe controlled spin-network, equipped with an AFM magnetic tip to generate gradient fields on nanometer length scales. (b) Extension of the multi-electron spectroscopy to extract inter-cell couplings, which can further lead to the structural analysis of the spin-network. By varying the $\pi$-pulse time on the second ($\tau$) we arrive at the Fourier components of the observed probe spin response. (c) Remote communication between two probes via a spin chain is shown, where the response on the probe spin $B$, $P_B (t) = \langle S^z_B \rangle$ varies by sweeping the frequency of the probe $A$. When $\omega_A = \omega_B$, the quantum state of probe $A$ is transferred to $B$. (d) The effect of the quantum critical point (QCP) in a transverse Ising chain on the state transfer between the two remote probes. At QCP corresponding to $\omega = J$, the fidelity drastically increases --- indicating enhanced long-range correlations in the chain.
 }
 \label{fig:fig4}
\end{figure*}

Indeed, using regular single-frequency spectroscopy as in Fig.\,\ref{fig:fig1}b, a $J$-coupling cannot be extracted as the network spins are unpolarized, which leads to averaging of all spectral spin-spin contributions. This results in observing only single peaks instead of doublets split by $J$. To tackle the problem in an analogous manner to regular spectroscopy, we utilize a third radio frequency (RF) to address different network spins simultaneously. Accordingly, it seems naturally conclusive to perform a decoupling sequence on the \textit{network} spins themselves while sweeping a third $\pi$-pulse and investigate its effect on the coherence created on the network spins (Fig.\,\ref{fig:fig3}a,b). This, in principle, amounts to addressing two external spins separately by two distinct RF frequencies.

However, since the coupling is small and the width of the RF pulses is non-negligible in our case, such a triple resonance scheme is sensitive to the inter-spin coupling of the network spins as
\begin{equation*}
 C_{\textrm{NV}}(\tau) \sim 1 - \prod_{k}^{M}{\sin^2\left( F^k \tau \right)} \cos \left(J^{\parallel} \tau \right) \cos \left(J^{\perp} \tau \right)
\end{equation*}
even when driving the network decoupling sequence at the Larmor frequency of the network spins and having a substantial overlap of the excitation pulses (see SI). Here the coupling between network spins is described by a parallel $J^{\parallel}$ and a perpendicular $J^{\perp}$ term (see Methods). We would like to stress that a proper choice of the excitation frequencies here can give access to the different elements of the dipolar coupling tensor. 

We apply the triple resonance scheme on the previously investigated NV-center probe environment. Here, the decoupling of the network spins is performed at their Larmor frequency, while sweeping the frequency of a third $\pi$-pulse. The resulting spectrum is shown in Fig.\,\ref{fig:fig3}a and shows two distinct peaks, with an effective coupling strength of $5.2 \pm 1.0$ MHz. To support these findings, we fix the second excitation frequency at the right peak and sweep the delay within the decoupling sequence, as we did to determine the probe-network coupling. Using this sequence the network-network coupling is probed and accordingly the result is a distinct oscillation, shown in Fig.\,\ref{fig:fig3}b, with a frequency of $5.8 \pm 1.0$ MHz, in excellent agreement with the previous measurement. Both measurements are supported by numerical simulations, reproducing the behavior for a $J$-coupled spin system probed by the NV spin sensor (see SI). Furthermore, we have performed control measurements, driving one of the two excitation channels far off-resonance to exclude any microwave artifacts, where we did not observe any coupling. In addition, we conducted measurements on a freshly cleaned sample with different electron spin systems, which did not show any hyperfine interaction (Fig.\,\ref{fig:fig1}e). These are most likely dangling bonds on the diamond surface and therefore not expected to exhibit any distinct dipolar coupling since their positions on the diamond surface are random. This was likewise not observed in our experiment.

\section*{Discussion and Outlook}
We have shown coupling of a single NV-center spin sensor and a few well-separated MTSSL network spins attached to synthetic peptides, accessible by non-invasive optical readout. Our triple resonance scheme allows us to furthermore extract the coupling between these network spins, which we specifically engineered via molecular chemistry. The demonstrated interface between a well-controllable probe spin and a coupled spin network opens a new venue for realizing a programmable spin network that could allow (i) the transfer of polarization (information) between remote probes, or perform long-distance gates acting as a quantum bus, (ii) for large scale simulation of solid-state spin Hamiltonians and (iii) for structural analysis of single large proteins using multi electron-electron spectroscopy at the nanoscale. With a continuous microwave driving of the probe one can polarize the network spins via the double resonance Hartman-Hahn techniques \cite{Hartmann1962, Atia2016}. Such polarized spin networks can narrow the linewidth of response on the probes, thus enhancing the fidelity of the network in achieving the targeted goals. In Fig.\,\ref{fig:fig4} we plot the expected response on the probe spins for the various protocols described above.

While cryogenic conditions enhance the ability to conduct long-term and highly reproducible electron spin spectroscopy meaurements of the targeted molecules, they also assist in implementing projective readout of the probe, adding more quantum features to the control of the network \cite{Greiner2016}. The measured network spin coherence time is close to the limit defined by the hydrogen nuclear spins in the MTSSL environment and this can be further improved by deuterating the molecule to which the electron spins are attached \cite{Zadrozny2015}. For generating tunable structures of the network one can perform lithography to define nanoscale features inhibiting growth in certain regions while binding molecules in others \cite{Gopinath2016}. Alternatively, self-assembly of peptides on specific surfaces can be controlled by sequencing and yield well-defined and regular two-dimensional structures \cite{Abb2016}. The triple resonance scheme can be generalized to access all elements of the coupling tensor and  the inter-scaffold couplings (see, for example, Fig.\,\ref{fig:fig4}b). While the electronic spins are used to form the network, the nuclear spins attached to them can act as a memory due to their long lifetime even at ambient conditions. Accessing and manipulating these spins could become helpful in understanding how information flows across a spin network and becomes available to multiple probes simultaneously (Fig.\,\ref{fig:fig4}c) and also as a probe for unveiling the quantum-critical behavior in a spin network (see Fig.\,\ref{fig:fig4}d).

\section*{Methods}
\subparagraph*{NV-DEER Spectroscopy}
The total Hamiltonian $H$ governing the network-spin dynamics is given by the sum of the sensor spin Hamiltonian $H_S$, the network spin Hamiltonian $H_{\sigma}$ and the sensor-network interaction Hamiltonian $H_{S\sigma}$\citep{supplement}:
\begin{equation}
\begin{split}
H =& H_S + H_\sigma + H_{S\sigma} \\
H_S =& DS^2_z +\gamma_SB S_z + \Omega(t)S_x \\
H_\sigma =& \sum_k \gamma_\sigma B \sigma^k_z + A_{zz}^k \sigma_z^k I_z^k + \omega(t)\sigma^k_x + \sum_{kl}\boldsymbol{\sigma}^k \cdot \mathbf {J}^{kl} \cdot \boldsymbol{\sigma}^l\\
H_{S\sigma} =& S_z\sum_k F^k \cdot \sigma^k_z
\end{split}
\label{eq1}
\end{equation}
In the above equation $B$ is the strength of the external magnetic field applied along the quantization axis of the NV-center probe spin, with spin operator $\boldsymbol{S} = (S_x, S_y, S_z)$, $S = 1$ and $D$ is its zero-field splitting. The dipolar coupling between the probe spin and the $k$-th network electron spins, with spin operator $\boldsymbol{\sigma}^k = (\sigma^k_x, \sigma^k_y, \sigma^k_z)$, $\sigma = 1/2$, is given by the scalar $F^k$ and the intra-spin coupling among the network spins is given by the dipolar coupling tensor $\mathbf{J}^{kl}$. The hyperfine coupling between the network electron spins and their respective nuclear spin environment is denoted by $A$ due to a nearest neighbor $^{14}\mathrm{N}$ nucleus with spin operator $\boldsymbol{I} = (I_x, I_y, I_z)$, $I = 1$. $\Omega(t)$ and $\omega(t)$ are the respective probe and network AC driving fields. When the Zeeman splitting of the networks spins along the field direction is much larger than their internal couplings one may neglect the non-secular terms, allowing to simplify the dipolar coupling between the spins to $\boldsymbol{\sigma}^k \cdot \mathbf {J}^{kl} \cdot\boldsymbol{\sigma}^l = J_{kl}^{\perp}\left(\sigma^k_{+} \sigma^l_{-} + \sigma^k_{-} \sigma^l_{+}\right) + J_{kl}^{\parallel} \sigma^k_z \sigma^l_z$. Similarly, due to the large difference in the Larmor frequencies between the electron and nuclear spins, the dominant contribution of the hyperfine interaction comes from  $A^k_{zz} \cos{(\theta_{k})}$, where $\theta_k$ is the relative angle between the external magnetic field and the orientation of the network spins.

The dipolar coupling between the network spins (that are deposited on the surface on the diamond) and the probe results in an additional phase evolution of the probe spin. To observe this phase one has to decouple the probe spin from the unwanted environmental phase noise. By flipping the probe spin at the midpoint of the decoupling sequence the random phase accumulated before and after the flip cancels. Thus, in order to keep the phase information of the network spins one needs to flip them along with the probe spin such that their contribution to the phase evolution of the probe is not canceled (see Fig.\,\ref{fig:fig1}b). This procedure, also known as double electron electron resonance (DEER) sequence, will lead to a contrast (i.e.\ an additional decoherence channel) in the readout of the probe spin as the microwave frequency is swept through the resonance of the network spins. The relative orientation between the molecular frame of MTSSL network spin and the magnetic field can then be found by rotating the magnetic field away from the quantization axis of the probe. As each MTSSL network spin can have a distinct orientation with respect to the applied field, knowledge of these satellite peaks might allow us to manipulate them individually.

\subparagraph*{Sample preparation}
For achieving close proximity between NV-centers and electron spins, we use an electronic grade diamond (Element 6), thinned down to a thickness of 30 $\upmu\mathrm{m}$ (Applied Diamond), in which we implant nitrogen ions at an energy of $5\ \mathrm{keV}$, yielding a mean implantation depth of $7\ \mathrm{nm}$ below the diamond surface \cite{Pezzagna2010}, in a dosage that allows to distinguish single NV-centers by standard confocal microscopy. Subsequently, we etch diamond nanopillars into the implanted membrane, to increase optical collection efficiency \cite{Momenzadeh2015}.

We select the nitroxide MTSSL spin label \cite{Berliner1982} due to the ease of the SDSL process and the well known spectral behavior \cite{Borbat2001}. The peptide H-AP$_{10}$CP$_{10}$CP$_{10}$-NH$_2$ was dissolved in 0.1 M Tris buffer (pH 7.8) with the addition of 30 mM tris(2-carboxyethyl)phosphine (TCEP) at a concentration of 1 mg/ml. The MTSSL spin label was dissolved in dimethylsulfoxid (0.1 mg/$\upmu$l). The peptide (114 nmol) was incubated with MTSSL (5 $\upmu$mol) at 4~$^\circ$C overnight. The free spin label was removed from the sample by a polyacrylamide desalting column, which was equilibrated with 0.1 M Tris buffer (pH 7.8) before the sample was applied to the column. The fractions which contain peptides were concentrated by a Pierce protein concentrator.  The spin-labeled peptide was dialyzed in ZelluTrans Mini Dialyzer against water to remove the buffer.  

After a standard three-acid (1:1:1 H$_2$SO$_4$:HNO$_3$:HClO$_4$) treatment on the diamond membrane, to clean and oxygen-terminate the diamond surface, we cover the diamond with $ \sim 3\ \upmu\mathrm{L}$ solution of $\mathrm{H}_2\mathrm{O}$, which contains $\sim 100\ \upmu\mathrm{M}$ peptides in a dropcasting fashion. The diamond is then situated in a nitrogen-rich atmosphere for around 1 hour, until all water has evaporated from the solution. This yields a surface coverage of $\geq 200\ \mathrm{nm}$ peptide crystallites. We mount the diamond membrane on a microwave-guide sample holder and span a $17\ \upmu \mathrm{m}$ gold wire across the surface in close proximity to the NV-centers of interest to supply high power microwave radiation for active manipulation of electronic spins. To conduct NV-center spin dependent fluorescence readout, we use a home-built confocal microscope, that contains both an ambient scanner unit, as well as a unit situated in a UHV environment, which can be cooled down to liquid helium temperatures \cite{SchaeferNolte2014}. All experiments were conducted under an external magnetic field of 100 \O rsted, aligned with the NV-axis.

\subparagraph*{EPR Spectroscopy}
To infer the size of the polyproline molecule, measurement of the dipolar coupling between electron spins on the cysteine sites in a commerical EPR spectrometer is performed, usually in a frozen deuterated buffer solution, and the distance is subsequently back-calculated. The derived distance distribution between the two cysteine sites on the above described polyproline is around $3.5 \pm 1.0$ nm \cite{Pannier2000, Qi2014} (see SI). The high error margin is caused here by the fact, that these measurements average over billions of molecules and the many resulting geometric conformations. It is important to note, that in this case it is actually the relative flexibility of the spin label tether which causes the variance in the distance distribution rather than the actual protein conformation itself. Assuming a polyproline helix II the distance between the two cysteine moieties should be around 3.1 nm, as determined by molecular dynamics simulations \cite{Schuler2005}.

\paragraph*{Materials}
The thiol-specific spin label MTSSL was obtained from Enzo Life Sciences (L\"orrach, Germany), TCEP was purchased  from Sigma Aldrich (Taufkirchen, Germany), Pierce protein concentrator PES 3K MWCO and polyacrylamide desalting columns 1.8 K MWCO  were obtained from ThermoFisher Scientific  and ZelluTrans/Roth Mini Dialyzer MD 300 MWCO 3500 were purchased from Carl Roth (Karlsruhe, Germany).

\paragraph*{Acknowledgements}
\noindent We thank A.\ Brunner, T.\ H\"aberle, D.\ Schmid-Lorch for fruitful discussions. We also thank M.\ Hagel for assistance in sample preparation and S.\ A.\ Momenzadeh for advice in diamond nanofabrication as well as S. Rauschenbach for help with peptide mass spectrometry. We acknowledge financial support from the European Union via grant no.\ 611143 (DIADEMS) and the DFG via SPP 1601 and Research Group 1493. F.F.d.O.\ acknowledges CNPq for the financial support through the project No. 204246/2013-0. A.F.\ acknowledges financial support from the Alexander von Humboldt Foundation.

\paragraph*{Author Contributions}
L.S., T.O., A.F. and J.W. conceived the experiment. 
L.S., K.X., B.K. and T.O. performed the NV measurements.
A.Z., M.A. and M.D. prepared the samples. M.A. performed the ESR spectrometer measurements.
F.F.d.O prepared the diamond membrane with NV-centers and A.F. fabricated nanopillars on it.
L.S., D.B.R.D. and A.F. analyzed the data and wrote the manuscript with input from all authors.
K.K., J.W. and A.F. supervised the project.

\bibliography{nppn_paper}

\end{document}